\documentstyle[11pt]{article}
\begin{document}
\title {\bf A note on the light velocity anisotropy}
\author{Bruno Preziosi \and
Dipartimento di Scienze Fisiche - Universit\`a Federico II di Napoli}
\date{}
\maketitle
{\it ABSTRACT}

{\it In the framework of linear transformations between inertial systems, 
there are no physical reasons for
assuming any anisotropy in the one-way velocity of light.}

{\it Pacs - 03.30 Special relativity 

}

 \section{Introduction}

The anisotropy in the microwave background$^{(1)}$ has suggested the existence 
of a preferred frame $\Sigma$ which sees an isotropic background and 
of a correspondent anisotropy in the
one-way velocity of light, when measured in our system $S$, which moves with
respect to $\Sigma$ at the velocity of about 377 Km/s.
This possibility has been exploited from the theoretical point of 
view$^{(2,3)}$;
many important and precise experiments have then been 
carried out with the purpose of detecting this anisotropy. No variation was 
observed at the level of $2 \times 10^{-13~(4)}$, $3\times 10^{-9~(5)}$ and 
$2 \times 10^{-15~(6)}$. 

We discuss here the generalized Lorentz transformations which have stimulated
these experiments.

\section {Local inertial systems}

The starting point of our discussion is the fact that 
the background radiation intensity appears to  be anisotropic to
an observer $O$,
at the origin of a reference frame $S$ in our "region" of the universe
$R$. It is quite reasonable to suppose, as we do, that the universe mass
distribution too is anisotropic from the point of view of $O$, while
it is seen isotropic by the observer $\Omega$ at the origin of a
preferred reference frame $\Sigma$.
We suppose moreover that all the reference frames we are speaking about
are settled in accordance with the following procedure:

$a)$ the origins are attached to bodies, which are "far enough" from other
celestial bodies and are well oriented with respect to identical very
far objects;

$b)$ identical conventional space and time standards are used in all the
reference systems.

The important point is that the absolute system $\Sigma$ and the  relative
frame $S$ of our region of the universe appreciate differences in the
radiation background, but cannot "locally" appreciate any force due to the
different mass distribution. The "region" $R$ behaves like the world 
inside the Einstein elevator; the Einstein equivalence principle ensures that,
if $\Omega$ and $O$ perform identical experiments in their respective
"regions", which are not influenced by the presence of local masses (Earth, 
Sun, ...), they obtain identical results.
 An immediate consequence is
that the inertia principle is valid for all the local inertial systems.

This concept is very clearly stated by Hans Reichenbach$^{(7)}$:

{\it It can be shown within the framework of Newton's theory that one can 
obtain local inertial systems by transforming away the gravitational field, 
although these systems are in a different state of motion provided that
the equivalence of inertial and gravitational mass is presupposed.
The gravitational field, which as such is still present, is compensated in 
these local systems by their acceleration relative to absolute space and 
the resulting inertial forces. According to Einstein, however, only these
local systems are the actual inertial systems. In them the field, which 
generally consists of a gravitational and an inertial component, is
transformed in such manner that the gravitational component disappears
and only the inertial component remains. The astronomical inertial systems
of Newton can at best be approximations which gradually change in the
neighborhood of stars. Only because distances in space are large compared 
to the masses of the stars, and because the stars have very low speeds, are 
astronomical inertial systems possible as approximations.}

The words "far enough" used in $a)$ express this approximation.

\section {Linear transformations}

If the transformations between these inertial systems are taken to be linear,
then the condition that any inertial motion is such for any local 
inertial system is implicity satisfied.
 Many authors have added, 
to this  hypothesis of linearity, other
hypotheses with the purpose of assuring consistency between the measures
of  different observers; many of them$^{(8-23)}$ have not postulated, a priori,
the invariance of the light speed. They conclude that these
transformations must be of Lorentz-type, characterized by a velocity
$\bf{c}$ which, in principle, may take different absolute value in the different
astronomical directions, but which is, for any fixed astronomical direction,
equal for all the local inertial systems. 

The consequences of the existence of a preferred reference frame $\Sigma$ 
have been
extensively discussed by Mansouri and Sexl$^{(2)}$; in particular they
 analyze the form the linear transformation from $\Sigma$ to an other
frame $S$ must take in order that the classical special relativity
experiments could be explained, without introducing {\it a priori} the
invariance of the speed of light. Moreover they compare the Einstein and
the transport syncronization procedure and derive the dependence of the
one-way velocity of light on its motion direction.

One of the most general transformations between the inertial systems $\Sigma$ 
and $S$, which
relatively move along the $x$-direction which connects the origins $\Omega$
and $O$, has been given by Robertson$^{(4)}$
in the form:
\begin{equation}
t'=a_{0}\left(t+\frac{va_{1}}{a_{0}c^{2}}x\right),~  
x'=a_{1}\left(x+\frac{va_{0}}{a_{1}}t\right),~  
y'=a_{2}y,~
z'=a_{2}z,~
\end{equation}
where $a_{0}$, $a_{1}$ and $a_{2}$ may depend on $v$.
This transformation, which is expressed in terms of the parameter $v$
and which reduces to the identity when $v=0$,
 is derived under the hypotheses that: $i)$ space is euclidean
for both $\Sigma$ and $S$ which use identical rods and clocks; $ii)$ in 
$\Sigma$ all clocks are syncronized and light moves with a speed $c$ which
is independent of direction and position; $iii)$ the one-way speed of
light in $S$ in the azymuth plane is direction independent.

Notice that the relative velocity of $S$ with respect to $\Sigma$
 is not given by $v$,
but is equal to $\tilde{v}\equiv v a_{0}/a_{1}$ ; analogously, the one-way 
velocity
of light is given by $\tilde{c}\equiv c a_{0}/a_{1}$. In terms of these
true velocities, equations (1) take the form
\begin{equation}
t'=a_{0}\left(t+\frac{\tilde{v}}{\tilde{c}^{2}}x\right),~
x'=a_{1}\left(x+\tilde{v}t\right),~
y'=a_{2}y,~
z'=a_{2}z.
\end{equation}

 The inverse transformation in the
$x$-direction is given by
\begin{equation}
x=\frac{1}{a_{1}(\tilde{v})(1-\frac{\tilde{v}^{2}}{\tilde{c}^{2}})}
\left(x'-\tilde{v}\frac{a_{1}(\tilde{v})}{a_{0}(\tilde{v})}t'\right). 
\end{equation}

Moreover, the velocity $\tilde{c}$ is seen by $S$ to take the value
\begin{equation}
\tilde{c}'=\frac{a_{1}(v)}{a_{0}(v)}\tilde{c}.
\end{equation}
In other words, the one-way velocity of light scales as the relative velocity
of the two frames. As a consequence, if $S$ and $S'$ are any two reference
frames which are moving (along the $x$-axis) with respect to $\Sigma$,
they appreciate different relative velocities and different same-way 
velocities of light which are in the same ratio. It follows that, if
the unit lenght of $S$ is such a way that $O$ and $\Omega$
appreciate the same relative speed, then $a_{0}=a_{1}$ and viceversa.

It is interesting to notice that, if the Robertson
hypotheses are released and one looks for the general 
one-dimensional linear transformations along the $x$-axis
 between the preferred frame $\Sigma$ 
and any $S$ which satisfies the condition that 

-not only $\Sigma$ and $S$  
appreciate the same relative speed but also any two frames $S$ and $S'$,
uniformly moving at finite velocities with respect to $\Sigma$, appreciate
the same finite speed, 

-then one fundamental velocity (which obviously must be identified with
the light speed) exists, the syncronization between two points of a
frame is in agreement with the Einstein rule and the transformations take 
the form of the first two of eqs. (2) with $a_{0}=a_{1}$ 
 (see references (6)-(20) and, in particular, (21)).

The transformation (2) is simply the product of the scale transformation which
scales the four dimensions respectively by the factors
\begin{equation}
a_{0}(\tilde{v})\sqrt{1-\frac{\tilde{v}^{2}}{\tilde{c}^{2}}},~~
a_{1}(\tilde{v})\sqrt{1-\frac{\tilde{v}^{2}}{\tilde{c}^{2}}},~~
a_{2}(\tilde{v}),~~
a_{2}(\tilde{v})
\end{equation}
by a standard Lorentz transformation.

The scale transformations, first considered by Galilei in {\it Discorsi
sovra due nuove Scienze}, were carefully analyzed by J. Fourier, H. von 
Helmoltz and
in particular by H. Poincar$\acute{e}$, who reached the conclusion that
a pure spatial deformation of the universe has no physical consequences; 
moreover, if the space and time coordinates of the universe are deformed 
in such manner that all the space-time coincidences are conserved, then
the universe remains unchanged.$^{(24)}$\\
If we consider the particular case of eqs. (2) in which $\tilde{v}=0$, we have
\begin{equation}
ds^{2}=dx'^{2}+dy'^{2}+dz'^{2}=a_{1}^{2}dx^{2}+a_{2}^{2}dy^{2}+a_{2}^{2}dz^{2}.
\end{equation}
This case is typical of a tetragonal crystal; the position of the second
nearest neighborhood is described in $S$ by the numbers $(1,1,0)$ and
in $S'$ by the components $(a_{1},a_{2},0)$. The distances remain the same.
It is worthwhile to notice that, in this case, if we scale the units of
$S'$ in such a way that they become equal, the times required by the light
to travel the reticular distances in the different directions remain
unchanged, but the light velocities are corresponding scaled; the
anisotropy of the tetragonal crystal remains untouched.

\section{Conclusions}
The general transformation (2) reduces to the Lorentz one if one requires
that  $\Omega$ and $O$ evaluate that the clock of the other observer is
going slower in the same ratio (or, equivalently, if each of them appreciate
an identical  contraction for the reciprocal rods). 

The effect of the scale transformations (5) must  be carefully analyzed.
For what concerns the lenghts in different directions, if there is no way 
for independent measures of lenghts and light velocities, in other words,
if unit lenghts are measured by fixing the light velocity (or viceversa),
there is no way of controlling the change in lenght of a rod with the 
orientation . The only thing to do is to use the Poincar$\acute{e}$
simplicity criterion and consider equal the lenghts of the rods and the
one-way speeds of light in the different directions.

In conclusion, isotropy in the one-way speed of light is a matter of
definition and the experiments conducted by J. Hall and coworkers at very 
sofisticated levels must be considered significant improvements of
some classical experiments in the frame of the special relativity.

There are however experiments which cannot be explained in the frame of
generalized linear Lorentz
transformation; this is the case when the time
required by the light to cover a poligonal is not in the ratio with
the time required to cover the apotheme, which is provided by the
Euclidean geometry. Notice that the
explanation is not found in supposing that the light speed along the
perimeter is different from the one along the apotheme (on the contrary,
from the point of view of a local observer the velocity of a photon
along its trajectory is always $c$$^(25)$), but in
requiring that the space structure is no more Euclidean.

Analogously, the experiments by Shapiro {\it et al.}$^{(26)}$ and
Reasenberg {\it et al.}$^{(27)}$ on the radar echo delay show that 
an e.m. wave travelling close to the Sun appears not only deviated, 
but also slightly retarded to an observer in a far flat region.
This phenomenon too cannot be explained by supposing a delay for the
light which comes from the Sun to the Earth, but must be settled in
the  general relativity frame.

\section{Acknowledgments}
Thanks are due to professors John L. Hall and Giuseppe Marmo for useful 
discussions.

\section{References}
1. G.F.Smoot, M.V. Gorenstein and R.A. Muller, {\it Phys. Rev. Lett.}
 {\bf 39}, 898 (1977)\\
2. R.M.Mansouri and R.U.Sexl, {\it J. Gen. Rel. Grav.} {\bf 8}, 497 (1977); 
{\bf 8},515 (1977); {\bf 8},809 (1977)\\
3. H.P.Robertson, {\it Rev. Mod. Phys.} {\bf 21}, 378 (1949); 
H.P.Robertson and 
T.W.Noonan, {\it Relativity and Cosmology}, Saunders, 
Philadelphia, (1968) p. 69\\
4. A. Brillet and J.L. Hall,  {\it Phys. Rev. Lett.}, {\bf 42}, 549 (1979)\\
5. E. Riis, L. Aaen Andersen, N. Bjerre, O.Poulsa, S.A.Lee, J. L. Hall,
{\it Phys. Rev. Lett.} {\bf 60}, 81, (1988)\\
6. Dieter Hils and J.L. Hall, {\it Phys. Rev. Lett.} {\bf 64}, 1697 (1990)\\
7. Hans Reichenbach, {\it Philosophie der Raum-Zeit-Lehre}
(Walter de Gruyter, Berlin and Leipzig, 1928) translated by M. Reichenbach
and J. Freund as
{\it The Philosophy of space and time} (Dover New York, N. Y. 1958)\\
8. W. von Ignatowsky, {\it Arch. Math. Phys. (Leipzig)} {\bf 17}, 1 (1910)\\
9. P. Frank and H. Rothe {\it Ann. Phys. Leipzig} {\bf 34}, 825 (1911)\\
10. F. Severi, {\it Rend. Regia Acc. Lincei}  1924 and
{\it Cinquant'anni di relativita',} Sansoni ed., Firenze, 1955\\
11. C. Cattaneo, {\it Rend. Acc. Lincei}, {\bf 24}, 256, (1958)\\
12. R. Weinstock, {\it Am. J. Phys.} {\bf 33}, 640, (1965)  \\
13. V. Mitvalsky, {\it Am. J. Phys.} {\bf 34}, 825 (1966)\\
14. L. J. Eisenberg, {\it Am. J. Phys.} {\bf 35}, 649 (1967)\\
15. V. Berzi and V. Gorini {\it J. Math. Phys.} 10, 1518 (1969)
16. A.R.Lee and T.M.Kalotas, {\it Am. J. Phys.} {\bf 43}, 5, 1975\\
17. Jean-Marc L\'evy-Leblond, {\it Am. J. Phys.} {\bf 44}, 271 (1976) and 
Riv. N.C. {\bf 7}, 187 (1977); Jean-Marc 
L\'evy-Leblond and Jean Pierre Provost, {\it Am. J. Phys.} {\bf 47}
, 12, (1979)\\
18. A.M. Srivastava, {\it Am. J. Phys.} {\bf 49}, 504 (1981)\\
19. N. D. Mermin,  {\it Am. J. Phys.} {\bf 52}, 119 (1984)\\
20. H. M. Schwartz, {\it Am. J. Phys.} {\bf 53}, 1007 (1985)\\
21. S. Singh, {\it Am. J. Phys.} {\bf 54}, 183 (1986)\\
22. A. W. Ross, {\it Am. J. Phys.} {\bf 55}, 174 (1987)\\  
23. B.Preziosi, {\it N. Cim.} {\bf 109 B}, 1331, (1994)\\ 
24. Moritz Schlick, {\it Raum und Zeit in der gegenw\"{a}rtigen Physik},
(Springer Verlag, Berlin, 1922)\\
25. I.R. Kenyon,{\it General relativity}, Oxford University Press, 1990,
p. 44\\
26. I.I. Shapiro {\it et al., Phys. Rev. Lett.} {\bf 26}, 1132 (1971)
27. R.D. Reasenberg {\it et al. Astr. J.} {\bf 234}, 1219 (1979)
\end{document}